\documentclass[journal]{IEEEtran}
\usepackage{graphicx}
\usepackage{multirow}
\usepackage{amsfonts}
\usepackage{amsmath}
\usepackage{stfloats}
\usepackage{color}
\usepackage{comment}
\usepackage{float}
\usepackage{flushend}
\usepackage{algorithm}
\usepackage{algorithmic}
\usepackage{amsthm}
\usepackage{amssymb}
\usepackage{epstopdf}
\usepackage{xcolor}
\usepackage{threeparttable}
\usepackage{booktabs}

\makeatletter
\renewcommand\normalsize{%
   \@setfontsize\normalsize\@xpt\@xiipt
   \abovedisplayskip 3\p@ \@plus3\p@ \@minus1\p@
   \abovedisplayshortskip 3\z@ \@plus3\p@
   \belowdisplayshortskip 5\p@ \@plus3\p@ \@minus1\p@
   \belowdisplayskip \abovedisplayskip
   \let\@listi\@listI}
\makeatother

\begin{document}

\title{Low-Complexity Block Coordinate Descend Based Multiuser Detection for Uplink Grant-Free NOMA}

\author{Pengyu Gao, Zilong Liu, \textit{Senior Member, IEEE}, Pei Xiao, \textit{Senior Member, IEEE}, Chuan Heng Foh, \textit{Senior Member, IEEE}, and Jing Zhang

\thanks{

P. Gao, P. Xiao and C. Foh are with 5G and 6G Innovation centre, Institute for Communication Systems (ICS) of University of Surrey, Guildford, GU2 7XH, UK (e-mail: p.gao@surrey.ac.uk; p.xiao@surrey.ac.uk; c.foh@surrey.ac.uk).

Z. Liu is with the School of Computer Science and Electronics Engineering, University of Essex, Colchester, CO4 3SQ, UK (e-mail: zilong.liu@essex.ac.uk).

J. Zhang is with Institute of space-based information systems, China Academy of Electronic and Information Technology (CAEIT), No. 11, Shuangyuan Road, Beijing 100041, China (e-mail: zhangjing-7@chinagci.com).
}
}

\maketitle

\begin{abstract}
Grant-free non-orthogonal multiple access (NOMA) scheme is considered as a promising candidate for the enabling of massive connectivity and reduced signalling overhead for Internet of Things (IoT) applications in massive machine-type communication (mMTC) networks. Exploiting the inherent nature of sporadic transmissions in the grant-free NOMA systems, compressed sensing based multiuser detection (CS-MUD) has been deemed as a powerful solution to user activity detection (UAD) and data detection (DD). In this paper, block coordinate descend (BCD) method is employed in CS-MUD to reduce the computational complexity. We propose two modified BCD based algorithms, called enhanced BCD (EBCD) and complexity reduction enhanced BCD (CR-EBCD), respectively. To be specific, by incorporating a novel candidate set pruning mechanism into the original BCD framework, our proposed EBCD algorithm achieves remarkable CS-MUD performance improvement. In addition, the proposed CR-EBCD algorithm further ameliorates the proposed EBCD by eliminating the redundant matrix multiplications during the iteration process. As a consequence, compared with the proposed EBCD algorithm, our proposed CR-EBCD algorithm enjoys two orders of magnitude complexity saving without any CS-MUD performance degradation, rendering it a viable solution for future mMTC scenarios. Extensive simulation results demonstrate the bound-approaching performance as well as ultra-low computational complexity.
\end{abstract}

\begin{IEEEkeywords}
Massive machine type communication (mMTC), grant-free, non-orthogonal multiple access (NOMA), compressed sensing (CS), block coordinate descend (BCD), multi-user detection (MUD), user activity detection (UAD), data detection (DD).
\end{IEEEkeywords}

\IEEEpeerreviewmaketitle

\section{Introduction}
\IEEEPARstart{M}assive machine-type communication (mMTC) is one of the three major use cases in the fifth generation (5G) mobile networks, which enables ubiquitous connectivity for various Internet-of-Things (IoT) applications \cite{mMTC-1}-\cite{mMTC-3}. Different from conventional human-centric communications, the design of an mMTC network needs to consider certain inherent features, such as small data size, sporadic traffic, low power consumption, etc. \cite{mMTC-4}, \cite{mMTC-5}.

Towards massive connectivity, the emerging non-orthogonal multiple acsess (NOMA) technique \cite{NOMA-1}-\cite{NOMA-3} is deemed as a promising candidate to accommodate a massive number of IoT devices. Unlike the classical orthogonal multiple acsess (OMA) techniques \cite{NOMA-1} where the number of supported devices is at most the number of available orthogonal resources, NOMA can concurrently support more communication links by sharing common resources in a non-orthogonal fashion. Thus, NOMA permits a higher spectrum efficiency compared with OMA, albeit at the expense of increased computational complexity at the receiver \cite{NOMA-2}. In addition, NOMA can be generally divided into two main categories, code-domain NOMA \cite{NOMA-1} and power-domain NOMA \cite{NOMA-3}, respectively. In this paper, we consider sequence based code-domain NOMA, where massive users shares the same time-frequency resources, but employ user-specific, non-orthogonal spreading sequences. Another challenging problem for mMTC is how each IoT device can rapidly access a channel resource. The long-term evolution (LTE) standard adopts a grant based uplink random access (RA) protocol with a four-phase handshaking procedure \cite{gf-1}. When IoT devices transmit small packets to the base station (BS) sporadically, however, the complicated signaling interactions between the BS and IoT devices result in potentially excessive signalling overhead and large access latency.

As a promising solution to the aforementioned problems, uplink grant-free NOMA communication is carried out based on non-orthogonal resource allocation in the code-domain, where each resource unit may be shared by more than one user \cite{gf-2}. Hence, it not only enables massive connectivity, but also allows active users transmit freely without waiting the grant of a communication network \cite{gf-2}-\cite{gf-5}. However, since the uplink channel resources allocated to the users in uplink grant-free NOMA are non-orthogonal, complicated multiple users detection may be needed at the BS. Note that in power-domain NOMA, unequal power allocation is adopted to separate different users. By contrast, code-domain NOMA allows equal power allocation to different users and the decoding does not rely on successive interference cancellation (SIC). Instead, in uplink grant-free NOMA, compressed sensing (CS) theory \cite{CS-1}-\cite{CS-4} is employed to jointly carry out user activity detection and data detection by exploiting the sporadic nature of mMTC transmissions. Specifically, CS is a powerful signal processing technique to reconstruct sparse signals from far fewer samples than that required by the Nyquist-Shannon sampling theorem \cite{CS-2}. So far, CS has been widely applied in wireless communications \cite{CS-5}, such as channel estimation \cite{CS-6} and spectrum sensing \cite{CS-7}.

By leveraging the feature of sporadic communications in mMTC, numerous CS-based multiuser detections (CS-MUD) have been studied in recent years \cite{CS-8}-\cite{fw-6}. Among these works, \cite{CS-8}-\cite{CS-10} only focused on single time-slot transmission model with no consideration of the temporal correlation. To further improve the performance of CS-MUD, frame-wise joint sparsity has been exploited, where the inactive and active users remain their individual transmission status during a whole frame \cite{fw-1}-\cite{fw-6}. In this case, CS-MUD at the BS can be generalized into a multiple measurement vector (MMV) problem. In \cite{fw-1}, a modified orthogonal matching pursuit (OMP) algorithm \cite{CS-3}, called iterative order recursive least square (IORLS), was developed for CS-MUD. To take full advantage of the frame-wise joint sparsity, Du \emph{et al}. in \cite{fw-2} integrated the block sparsity structure into the classical subspace pursuit (SP) algorithm \cite{CS-4} and proposed the threshold-aided block sparsity adaptive subspace pursuit (TA-BSASP) that can approach the oracle performance with no need of the user sparsity level as the prior knowledge. Furthermore, subspace matching pursuit based symbol correcting mechanism (SMP-SCM) in \cite{fw-3} has shown improved symbol error rate (SER) over oracle least square (LS) \cite{gf-3} at the expense of computational cost. However, a salient drawback of these greedy-based algorithms is the computationally prohibitive operations of the matrix multiplications and inversions, especially for large-scale problems in mMTC scenarios.
In parallel with greedy-based algorithms, many existing CS-MUD schemes have been developed based on the Bayesian approach \cite{fw-4}-\cite{fw-6}. The authors in \cite{fw-4} utilized approximate message passing (AMP) and expectation maximization (EM) to carry out MUD with the aid of the prior information of the transmitted symbols. Moreover, \cite{fw-5} introduced a maximum a posteriori probability (MAP) based active user detector by exchanging the extrinsic information. Mei \emph{et al}. in \cite{fw-6} developed an orthogonal approximate message passing (OAMP) based algorithm by leveraging the prior knowledge of the discrete constellation symbols and structured transmission sparsity. However, the Bayesian-based methods may suffer from sophisticated probabilistic iterations and the inaccurate Gaussian approximation at the receiver side, precluding them from efficient practical implementations. 

This paper presents the design of a novel CS-MUD detector which enjoys an excellent MUD performance and low computational complexity. For this purpose, we first transform the MUD problem into a group least absolute shrinkage and selection operator (LASSO) problem and then address it globally by the block coordinate descent (BCD) based algorithms \cite{BCD-1}. Particularly, in the conventional BCD framework, the formulated non-convex sparse recovery problem can be decomposed into a sequence of small-scale subproblems after exploiting LASSO based regularization \cite{BCD-2}. Subsequently, the variables in each subproblem can be optimized according to the closed-form solution sequentially with variables from other subproblems keeping fixed. In uplink grant-free NOMA systems, the updating rules of the BCD method indicates that all signals, regardless of active or inactive users, are calculated and then further reused to update others signals. In particular, we observe that estimating the signals of inactive users indeed is to reconstruct the noise. In this case, severe inter-user interference may be inevitable, making active users indistinguishable. Motivated by this, we first develop an enhanced BCD (EBCD) algorithm which can identify a majority of inactive users in advance, thus helping significantly reducing inter-user interference. Next, we further improve the proposed EBCD algorithm for computational complexity saving. The resultant algorithm is called complexity-reduction enhanced BCD (CR-EBCD), which shares the same MUD performance as that of EBCD, whilst having significantly reduced computational complexity. Without the need of knowing the user sparsity level as priori information, our proposed two BCD-based algorithms enable the identification of active users adaptively upon the energy-based threshold. Specifically, our major contributions are summarized as follows:

\begin{itemize}
\item{\textbf{\textit{Proposed EBCD Algorithm}}: In the proposed EBCD algorithm, we introduce a candidate set pruning mechanism to improve the CS-MUD performance. More explicitly, different from the original BCD method where all potential users are involved in data updating in each iteration, our proposed EBCD algorithm discards inactive users according to the signal power at the early iterations. Furthermore, the corresponding signals of those discarded users are set to zeros for direct interference mitigation. Therefore, by utilizing the pruning candidate, the CS-MUD performance is significantly improved compared to that of the original BCD algorithm. In addition, with the decreasing number of users in the candidate set, the computational overhead is also reduced.

}
\end{itemize}

\begin{itemize}
\item{\textbf{\textit{Proposed CR-EBCD Algorithm}}: As a further extension of the proposed EBCD, our proposed CR-EBCD dramatically simplifies the iteration process by removing the redundant matrix multiplications. Consequently, the calculations in the proposed CR-EBCD are reduced into the simple calculation of vector multiplications, which avoids complicated large-scale matrix multiplications. Both algorithms are able to achieve the same bound-approaching performance with the aid of the candidate set pruning mechanism, but the proposed CR-EBCD can also enjoy about two orders of magnitude computational complexity reduction. Therefore, our proposed CR-EBCD strikes an excellent tradeoff between the detection performance and computational complexity.
}
\end{itemize}

It is important to point out that the core idea of the previous studies \cite{LowCo-1}-\cite{LowCo-2} focusing on low-complexity CS-MUD design in uplink grant-free NOMA systems is to substitute the complicated matrix inversion operations by simple gradient descend. However, since they follow greedy-based algorithms, calculating the correlation at each iteration inevitably leads to high computational cost due to the large-scale matrix multiplications. By contrast, the simple operations between vectors in our proposed BCD-based algorithms not only avoid the matrix inversion operations, but require no matrix multiplications.

The rest of the paper is organized as follows. In Section II, the system model and problem formulation are presented. Next, the original BCD algorithm, our proposed EBCD and CR-EBCD algorithms are detailed sequentially in Section III. Then, Section IV provides the computational complexity analysis. Simulation results are given in Section V to verify the effectiveness of our proposed algorithms. Finally, Section VI concludes the paper.

\emph{Notations:} Boldface capital and lowercase symbols represent matrices and column vectors, respectively. The ${\left(  \cdot  \right)^T}$, ${\left(  \cdot  \right)^H}$ and ${\left( \cdot  \right)^{-1}}$ operations represent transpose, Hermitian transpose and inverse, respectively. ${\left(  \cdot  \right)^\dag }$ denotes pesudo-inverse operation. $\left\|\cdot \right\|_p$ is the $l_p$-norm operation. Besides, $\text{diag}\left( \cdot \right)$ denotes a diagonal matrix with the elements on its diagonal. $\odot$ stands for the Hadamard product. ${{\bf{I}}_{N}}$ means the identity matrix of size $N \times N$. Additionally, $\max \left( {\Omega,k} \right)$ represents selecting the $k$ largest elements from the set $\Omega$. Furthermore, for a matrix $\mathbf Z$, $\mathbf Z\left[ \Lambda  \right]$ refers to the sub-matrix of $\mathbf Z$ that only contains those columns indexed by the set $\Lambda$. ${\left\{ {1,2, \cdot  \cdot  \cdot ,K} \right\}\backslash {\Lambda}}$ represents that the set contains ${\left\{ {1,2, \cdot  \cdot  \cdot ,K} \right\}}$, but excludes the elements in set $\Lambda$. Finally, ${\cal C}{\cal N}\left( {0,{\sigma ^2}}\right)$ represents the complex Gaussian distribution with zero mean and variance $\sigma^2$.

\section{System Model}
Consider an uplink grant-free NOMA system, where users are static (or near static) and a BS communicates with a total of fixed $K$ users \cite{fw-11}, as illustrated in Fig. 1. Without loss of generality, we assume that the BS and all users are equipped with a single antenna.

\begin{figure}[t]
\centering{\includegraphics[width=80mm]{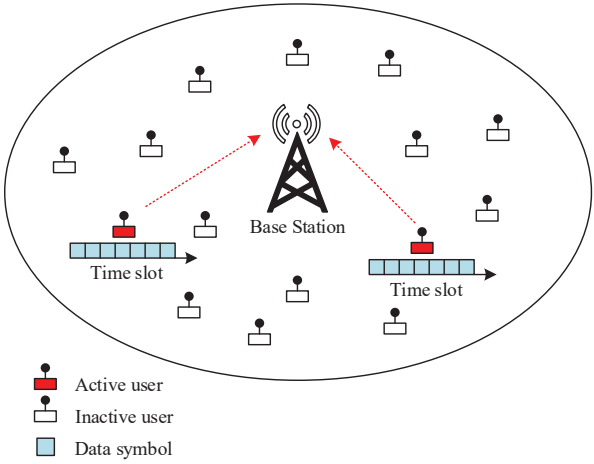}}
\caption{System model of an uplink grant-free NOMA system in one transmission frame. Due to sporadic nature of mMTC transmissions, only a small fraction of users are active while others keep silent. Additionally, the transmission status of active or inactive users remains unchanged during an entire frame.}
\end{figure}

In an arbitrary time interval, although the total number of users may be very large, it is reasonable to assume that only a small fraction of users are active due to the sporadic traffic nature of the mMTC transmissions. For active user $k$, the transmitted symbol $x_k$ is selected from a complex-constellation set $\chi $. Meanwhile, if user $k$ is silent, $x_k$ can be regarded as zero. Hence, the augmented complex-constellation set $\widetilde \chi$ can be represented by $\widetilde \chi  \buildrel \Delta \over = \left\{ {\chi  \cup 0} \right\}$. Then, data symbols are spread onto $N$ subcarriers by user-specific spreading sequences. Subsequently, the signals from all the active users are superposed over the same $N$ orthogonal resources at the receiver side. Particularly, to meet the demand of massive connectivity in mMTC scenarios, code-domain NOMA technique is considered and non-orthogonal spreading sequences are assigned to each user, indicating that the length of each spreading sequence is less than the total number of users in the systems, i.e., $N < K$.

At the BS, the frequency-domain received signal ${\bf{y}} \in {\mathbb{C}}^{N \times 1}$ can be expressed as
\begin{equation}
{\bf{y}} = \sum\limits_{k = 1}^K {\left( {{{\bf{h}}_k} \odot {{\bf{s}}_k}} \right){x_k}}  + {\bf{v}},
\end{equation}
where ${{\bf{h}}_k} = {\left[ {{h_{1,k,}},{h_{2,k,}}, \cdots ,{h_{N,k}}} \right]^T} \in {\mathbb{C}}^{N \times 1}$ is the channel coefficients between user $k$ and the BS, whose elements follow the independent complex Gaussian distributions ${h_{n,k}} \sim {\cal C}{\cal N}\left( {0,1} \right)$. Here, as many previous works \cite{fw-1}-\cite{fw-6} did, perfect multi-user synchronization\footnotemark and reliable channel estimation\footnotemark[2] are assumed at the BS. ${{\bf{s}}_k} = {\left[ {{s_{1,k}},{s_{2,k}}, \cdots ,{s_{N,k}}} \right]^T} \in {\mathbb{C}}^{N \times 1}$ is the unique spreading sequence of user $k$. In addition, ${x_k}$ represents the transmitted constellation symbol from active user $k$ to the BS. For the transmission power control, we have $E\left[ {{{\left| {{x_k}} \right|}^2}} \right] = 1$. Moreover, ${{\bf{v}}}\sim{\cal C}{\cal N}\left( {0,\sigma^2{{\bf{I}}_{N}}} \right)$ is the additive white Gaussian noise (AWGN).

\footnotetext[1]{This can be achieved by sending beacon signals periodically from the BS \cite{fw-7}. At the same time, all the uplink users constantly adjust their synchronization points by aligning with that of the beacon signals.}

\footnotetext[2]{We assume that channel estimation is carried out at the BS for all users (including those inactive ones) through a dedicated control channel. Furthermore, since we consider the mMTC scenario, users in the systems are mostly static with fixed locations \cite{mMTC-5}. Thus, it is reasonable that the channel information obtained by the BS can be utilized during a long data transmission period.}

Accordingly, (1) can be equivalently rewritten as
\begin{equation}
{\bf{y}} = \sum\limits_{k = 1}^K {\text{diag}\left( {{{\bf{h}}_k}} \right){{\bf{s}}_k}{x_k} + {\bf{v}} = {\bf{Gx}} + {\bf{v}}},
\end{equation}
where ${\bf{G}} \in {\mathbb{C}}^{N \times K}$ denotes the equivalent channel matrix with elements ${g_{n,k}} = {h_{n,k}}{s_{n,k}}$. The transmitted signals of all $K$ users are represented as
${\bf{x}} = {\left[ {{x_1},{x_2}, \cdots ,{x_K}} \right]^T} \in {\mathbb{C}}^{K \times 1}$.

In this paper, we consider the frame-wise joint sparsity model, where the transmission status of active and inactive users remain unchanged during an entire frame, as shown in Fig. 1. Assuming that a data frame contains $J$ continuous time slots, the frame-wise joint sparsity can be formulated as
\begin{equation}
{\mathop{\rm supp}\nolimits} ( {{{\bf{x}}^{\left( 1 \right)}}} ) =  \cdot  \cdot  \cdot  = {\mathop{\rm supp}\nolimits} ( {{{\bf{x}}^{\left( J \right)}}} ) = \Gamma,
\end{equation}
where ${\mathop{\rm supp}\nolimits} \left( {{{\bf{x}}^{\left( j \right)}}} \right) = \{ {k  \left| {x_k^{\left( j \right)} \ne 0,1 \le k \le K} \right.} \}$, the support $\Gamma $ can be described as the candidate set, which indicates the set of active users. $\left| \Gamma \right|$ denotes the total number of active users. Furthermore, the channel remains unchanged during a whole frame, i.e. ${{\bf{G}}^{\left( 1 \right)}} =  \cdot  \cdot  \cdot  = {{\bf{G}}^{\left( J \right)}} = {\bf{G}}$. Due to the frame-wise joint sparsity and static channel, the received signals in frequency-domain during a certain frame can be rewritten as
\begin{equation}
{\bf{Y}}{\rm{ = }}{\bf{GX}} + {\bf{V}}{\rm{ = }}\sum\limits_{k = 1}^K {{{\bf{g}}_k}} {\bf{x}}_k^T + {\bf{V}},
\end{equation}
where ${\bf{Y}} = \left[ {{{\bf{y}}^{\left( 1 \right)}},{{\bf{y}}^{\left( 2 \right)}}, \cdots ,{{\bf{y}}^{\left( J \right)}}} \right] \in {\mathbb{C}}^{N \times J}$, ${\bf{X}} = \left[ {{{\bf{x}}^{\left( 1 \right)}},{{\bf{x}}^{\left( 2 \right)}}, \cdots ,{{\bf{x}}^{\left( J \right)}}} \right] \in {\mathbb{C}}^{K \times J}$ and ${\bf{V}} = $$\left[ {{{\bf{v}}^{\left( 1 \right)}},{{\bf{v}}^{\left( 2 \right)}}, \cdots ,{{\bf{v}}^{\left( J \right)}}} \right] \in {\mathbb{C}}^{N \times J}$, respectively. In addition, ${{\bf{g}}_k}\in {\mathbb{C}}^{N \times 1}$ denotes the $k$-th column of ${\bf{G}}$, which represents the equivalent channel coefficients of user $k$. ${\bf{x}}_k^T\in {\mathbb{C}}^{1 \times J}$ denotes the $k$-th row of ${\bf{X}}$, representing the data symbols of user $k$ transmitting over the consecutive $J$ time slots.

The BS aims at recovering the transmitted signal ${\bf{X}}$ based on the received signal ${\bf{Y}}$. It is noted that each column in ${\bf{X}}$ is a sparse vector due to the sporadic transmission nature of mMTC data traffic. Moreover, with the frame-wise joint sparsity transmission model, ${\bf{X}}$ is a sparse matrix consisting of many zero rows. Thus, the sparse construction problem in (3) can be regarded as a multiple measurement vector (MMV) based CS problem.

\section{The Proposed BCD based Algorithms}
In this section, we first introduce the original BCD algorithm for uplink grant-free NOMA systems. Subsequently, we present our proposed EBCD algorithm which is shown to achieve a near-oracle MUD performance and enjoy low computational complexity, thanks to the candidate set pruning mechanism. Finally, our proposed CR-EBCD algorithm for further computational complexity reduction is described.

\subsection{The Original BCD Algorithm}
The objective of the BS is to solve the sparse recovery problem in (4), which can be equivalently formulated into an optimization problem as follows:
\begin{equation}
\begin{array}{l}
\mathop {\min }\limits_{\left\{ {{{\bf{x}}_k}} \right\}_{k = 1}^K} \left\| {{\bf{Y}} - \sum\limits_{k = 1}^K {{{\bf{g}}_k}} {\bf{x}}_k^T} \right\|_2^2\\
s.t{\kern 1pt} {\kern 1pt} {\kern 1pt} {\kern 1pt} {\kern 1pt} {\kern 1pt} \sum\limits_{k = 1}^K {{{\left\| {{\bf{x}}_k^T} \right\|}_0}}  \le K_{\max }^{active}
\end{array},
\end{equation}
where ${{{\bf{g}}_k}}$ is the $k$-th column of ${\bf{G}}$ and $K_{\max }^{active}$ denotes the maximum number of active users in the system. Since the optimization problem in (5) is NP-hard due to the non-convex sparsity constraint \cite{CS-11}, it is thus computationally intractable. A straightforward solution is to carry out exhaustive search for the maximum likelihood (ML) detection \cite{fw-3}. However, this involves a total number of  ${\left( {M + 1} \right)^K}$ options, considering all possible active user combinations and the constellation symbol combinations, where $M$ is the modulation cardinality. For instance, when quadrature phase shift keying (QPSK) modulation is adopted, $M = 4$. As such, the computational complexity of the exhaustive search in ML method is prohibitively high, which precludes it from practical implementation especially for large value of $K$ in the future mMTC scenarios.

To tackle this bottleneck, the group LASSO technique can be employed, which provides the convex relaxation of such discontinuous non-convex constraint in (5). Based on the group LASSO framework, (5) can be approximated in an equivalent penalized form as
\begin{equation}
\mathop {\min }\limits_{\left\{ {{{\bf{x}}_k}} \right\}_{k = 1}^K} \frac{1}{2}\left\| {{\bf{Y}} - \sum\limits_{k = 1}^K {{{\bf{g}}_k}} {\bf{x}}_k^T} \right\|_2^2 + \lambda {\kern 1pt} \sum\limits_{k = 1}^K {\left\| {{{\bf{x}}_k}} \right\|_2^2},
\end{equation}
where $\lambda>0$ is the penalty parameter for sparsity constraint, also called the tuning factor, which is usually selected by trial-and-error. More importantly, the choice of $\lambda$ should be meticulous, since it is critical for obtaining the optimal solution towards the above problem. Specifically, given a large value of $\lambda$, the solution of (6) could be sufficiently sparse. On the other hand, if the value of $\lambda$ approaches to zero, the sparsity of the solution tends to be $K$. It is noted that (6) is a convex optimization problem and the optimal objective is to identify non-zero rows in ${\bf{X}}$. In the following, a series of the BCD based algorithms are applied, which can address this group LASSO problem efficiently in a globally optimal way.

In the BCD framework, (6) can be decoupled into a sequence of $K$ small subproblems. In particular, each group of variables
${{{\bf{x}}_k}}$ in certain subproblem will be optimized one by one while keeping other group of variables ${{\bf{x}}_l}$ ($l \in \left\{ {1,2,...,K} \right\}\backslash k$) fixed. To be specific, the $k$-th subproblem degenerating from (6) can be expressed as
\begin{equation}
\mathop {\min }\limits_{{{\bf{x}}_k}} \frac{1}{2}\left\| {{{\bf{R}}_k} - {{\bf{g}}_k}{\bf{x}}_k^T} \right\|_2^2 + \lambda {\kern 1pt} \left\| {{{\bf{x}}_k}} \right\|_2^2,
\end{equation}
where ${{\bf{R}}_k}{\rm{ = }}{\bf{Y}}{\rm{ - }}\sum\limits_{l = 1,l \ne k}^K {{{\bf{g}}_l}} {\bf{x}}_l^T$. Here, we define a convex function as follows
\begin{equation}
f\left( {{{\bf{x}}_k}} \right) \buildrel \Delta \over = \frac{1}{2}\left\| {{{\bf{R}}_k} - {{\bf{g}}_k}{\bf{x}}_k^T} \right\|_2^2 + \lambda {\kern 1pt} \left\| {{{\bf{x}}_k}} \right\|_2^2.
\end{equation}
According to (8), we can obtain the optimal solution of ${\bf{x}}_k$ in a closed form by setting the first-order derivation of $f\left( {{{\bf{x}}_k}} \right)$ to zero. Hence, the optimal solution ${\bf{x}}_k^*$  can be obtained by
\begin{equation}
{\left( {{\bf{x}}_k^ * } \right)^T} = \frac{{{\bf{g}}_k^H{{\bf{R}}_k}}}{{{\bf{g}}_k^H{{\bf{g}}_k} + {\lambda _0}}},
\end{equation}
where $\lambda _0 = 2\lambda$. Updating the optimal solution following the (8) and (9), the results will reach convergence after sufficient iterations.

The main procedure of the original BCD algorithm is stated as \textbf{Algorithm 1}. As can be seen, the data updating process happens in Step 2. In the sequel, the energy-based threshold is utilized to identify the non-zero positions in Step 3. Particularly, ${V_{th}}$ is relative to the signal-to-noise ratio, which can be selected empirically \cite{gf-3}. After total $T$ iterations, once the signal power of user $k$ exceeds the threshold, i.e., $\left\| {{\hat{\mathbf{x}}_k}} \right\|_2^2 > {V_{th}}$, user $k$ will be regarded as active user and the index $k$ will be stored in the final candidate set $\Gamma$. Finally, for a more balanced SER performance and computational complexity trade-off, LS method is exploited to recover the transmitted signal with the aid of the estimated candidate set in Step 4, rather than directly using the results from the previous iteration process. The reason is that the interference introduced by inactive users have adverse influences on estimating the transmitted signals. Conversely, as for the LS method, we only use the estimated support to reconstruct the signals for a more reliable detection results.

\begin{algorithm}[htb]
\caption{The Original BCD Algorithm.}
\label{algorithm 1}
\begin{algorithmic}[1] 
\REQUIRE ~~ \\ 
The received signal: $\mathbf{Y}$;\\
Equivalent channel matrix: $\mathbf{G}$; \\
The maximum number of iterations: $T$; \\
The equivalent regularization parameter: ${\lambda _0};$\\
The threshold parameter: ${{V}_{th}}$; \\
\ENSURE ~~ \\ 
Reconstructed sparse signals: $\hat {\bf{X}}{\rm{ = [}}{\hat {\bf{x}}_1^T}{\rm{,}}{\hat {\bf{x}}_2^T}{\rm{,}}...{\rm{,}}{\hat {\bf{x}}_k^T}{\rm{]}}$.\\

\textbf{$\bullet$ Step 1} (\emph{Initialization})\\

\STATE (Iteration initialization): the iterative index $t=1$, the support set ${\Gamma}=\emptyset$, the initial transmitted signal ${\hat{\mathbf{x}}_k} = {\bf{0}},\forall k = 1,2,...,K$.

\textbf{$\bullet$ Step 2} (\emph{Iteration})\\

\textbf{for} {$t = 1,2,...,T$} \textbf{do}\\
\quad \textbf{for} $k = 1,2,...,K$ \textbf{do}\\

\STATE \quad \quad Compute ${{\bf{R}}_k^{\left( t \right)}}$.

\STATE \quad \quad Update ${\hat{\mathbf{x}}}_{k,\left( t \right)}^T$ using (9).

\quad \textbf{end for}\\
\textbf{end for}\\
\textbf{$\bullet$ Step 3} (\emph{Obtain The Estimated Candidate Set})\\

\textbf{for} {$k = 1,2,...,K$} \textbf{do}\\
\quad\textbf{if} $\left\| {{\hat{\mathbf{x}}_k}} \right\|_2^2 > {V_{th}}$ \textbf{then}
\STATE \quad \quad $\Gamma \leftarrow \Gamma  \cup \left\{ k \right\}$. \\
\quad\textbf{end if}\\
\textbf{end for}\\

\textbf{$\bullet$ Step 4} (\emph{Obtain Data Symbols})\\
\STATE
$\hat {\bf{X}}\left[ \Gamma  \right] = {\left( {{\bf{G}}\left[ \Gamma  \right]} \right)^\dag }{\bf{Y}}$,
$\hat {\bf{X}}\left[ {\left\{ {1,2,...,K} \right\}\backslash \Gamma } \right] = 0$.\\

\end{algorithmic}

\end{algorithm}

\subsection{The Proposed EBCD Algorithm}
Note that the original BCD method is a global optimization algorithm, where the variables from ${\bf{x}}_1^T$ to ${\bf{x}}_K^T$ are updated sequentially at each iteration. However, since the majority of elements in ${\bf{x}}_k^T$ are zeros due to sporadic transmissions, the updating process of those zero positions can be regarded as computational wastage. Moreover, updating the signals of inactive users indeed is to reconstruct the noise. In this case, signals from inactive users are bound to deteriorate the detection performance since they are used to estimate other users' signals at each iteration when calculating ${{\bf{R}}_k}$. Motivated by this, we develop a novel EBCD algorithm, which can identify the majority of inactive users in advance, thanks to the design of the candidate set pruning mechanism. To be specific, in our proposed EBCD algorithm, a majority of inactive users can obtain their own optimal solutions (i.e., ${\bf{0}}$) in advance instead of running the iterative algorithm until convergence, which is beneficial for the identification of active users due to the interference mitigation. Furthermore, since a great number of inactive users can be discarded halfway through the iteration process, our proposed EBCD algorithm also enjoys a considerable computational complexity reduction.

For describing our proposed EBCD algorithm explicitly, we firstly define a parameter $\Omega^{\left( {t} \right)}$ to indicate the candidate set at the $t$-th iteration and $\Omega_i^{\left( {t} \right)}$ denotes the $i$-th element in $\Omega^{\left( {t} \right)}$. Correspondingly, ${K_a^{\left( {t} \right)}}$ is defined to measure the scale of $\Omega^{\left( {t} \right)}$. In particular, the candidate set $\Omega$ in the original BCD method consists of all the potential users in the system, i.e. $\Omega = \left\{ {1,2,...,K} \right\}$. In other words, if ${K_a^{\left( {t} \right)}}$ always equals to $K$ at each iteration, our proposed EBCD algorithm boils down to the original one. Therefore, at each iteration, only ${K_a^{\left( {t} \right)}}$ elements in $\Omega^{\left( {t} \right)}$ need to be updated according to (9).

With respect to the candidate set pruning, the signal power is employed as the criteria, since a higher value of ${\left\| {{{\widehat {\bf{x}}}_{\Omega_i^{\left( {t} \right)}}}} \right\|_2^2}$ signifies that the user with the corresponding index $\Omega_i^{\left( {t} \right)}$ is more likely to be active. According to this principle, ${\psi ^{\left( t \right)}}$ users holding the lower signal power are deemed as inactive users and will be discarded from the current candidate set $\Omega^{\left( {t} \right)}$. Therefore, the scale of the new candidate set ${\Omega ^{\left( {t + 1} \right)}}$ is updated by 
\begin{equation}
{K_a^{\left( {t+1} \right)}} = {K_a^{\left( {t} \right)}} - {\psi ^{\left( t \right)}}.
\end{equation}
Then, we can obtain a new candidate set ${\Omega ^{\left( {t + 1} \right)}}$ by pruning the old candidate set ${\Omega ^{\left( {t} \right)}}$ as follows:
\begin{equation}
    {\Omega ^{\left( {t + 1} \right)}}{\rm{ = }}\max \left( {\left\| {{{\widehat {\bf{x}}}_{\Omega_i^{\left( {t} \right)}}}} \right\|_2^2,K_a^{\left( {t + 1} \right)}} \right), {\kern 1pt} {\kern 1pt} {\kern 1pt} {\kern 1pt} {\Omega_i^{\left( {t} \right)}} \in {\Omega ^{\left( t \right)}}.
\end{equation}
The key idea of (11) is to select the largest $K_a^{\left( {t + 1} \right)}$ elements from ${\Omega ^{\left( {t} \right)}}$  to constitute the new candidate set ${\Omega ^{\left( {t + 1} \right)}}$. Meanwhile, the remaining elements from ${\Omega ^{\left( {t} \right)}}/{\Omega ^{\left( t + 1\right)}}$ are set to zeros, i.e. ${\widehat {\bf{x}}_{{\Omega ^{\left( {t} \right)}}/{\Omega ^{\left( t + 1\right)}}}} = {\bf{0}}$. Actually, zeros are the optimal solutions for the inactive users. Hence, (10) and (11) can not only prune the candidate set for the next iteration, but also provide the optimal solutions for ${\psi ^{\left( t \right)}}$ discarded users. That is, in the subsequent iterations, the amount of computations will be decreased while the calculation results of ${\bf{R}}_k$ will also be more accurate for estimating the active users' signals. It is important to note that after $T_b$ iterations, the final candidate set keeps invariant and only contains ${\overline K _a}$ users, which can be calculated by
\begin{equation}
{\overline K _a} = {K - \sum\limits_{t = 1}^{{T_b}} {{\psi ^{\left( t \right)}}}}.
\end{equation}
To avoid discarding the non-zero candidates, we set ${\overline K _a} \ge 0.2K$, considering the sporadic transmission in the practical scenarios\footnotemark[3].

It is noted that the procedure of our proposed EBCD algorithm will be exhibited together with our proposed CR-EBCD algorithm in order to avoid too much repetition. In the sequel, it will be explained detailedly.
\footnotetext[3]{Under normal circumstances, the number of active users does not exceed 10${\rm{\% }}$ of the total number of potential users even in the busy hours \cite{gf-6}, \cite{gf-7}.}

\begin{algorithm}[!h]
\caption{Proposed CR-EBCD Algorithm.}
\label{algorithm 3}
\begin{algorithmic}[1] 
\REQUIRE ~~ \\ 
The received signal: $\mathbf{Y}$;\\
Equivalent channel matrix: $\mathbf{G}$; \\
The maximum number of iterations: $T$; \\
The equivalent regularization parameter: ${\lambda _0};$\\
The threshold parameter: ${{V}_{th}}$; \\
The parameters for pruning the candidate set: \\$\Psi {\rm{ = }}\left\{ {{\psi ^{\left( 1 \right)}},{\psi ^{\left( 2 \right)}},...,{\psi ^{\left( {{T_b}} \right)}}} \right\}$;\\
\ENSURE ~~ \\ 
Reconstructed sparse signals: $\hat {\bf{X}}{\rm{ = [}}{\hat {\bf{x}}_1^T}{\rm{,}}{\hat {\bf{x}}_2^T}{\rm{,}}...{\rm{,}}{\hat {\bf{x}}_k^T}{\rm{]}}$.\\

\textbf{$\bullet$ Step 1} (\emph{Initialization})\\

\STATE (Iteration initialization): The iterative index $t=1$, the support set ${\Gamma}=\emptyset$, the initial candidate set $\Omega^{\left( {1} \right)} {\rm{ = }}\left\{ {1,2,...,K} \right\}$, the initial scale of the candidate set ${{K_a^{\left( {1} \right)}}} = K$, the transmitted signals ${\hat{\mathbf{x}}_{k,\left( 1 \right)}^T} = {\bf{0}},\forall k = 1,2,...,K$.\\
\textbf{$\bullet$ Step 2} (\emph{Iteration})\\

\textbf{for} {$t = 1,2,...,T$} \textbf{do}\\
\quad \textbf{for} $i = 1,...,{K_a^{\left( {t} \right)}}$ \textbf{do}\\
\quad \quad\textbf{if} $i = 1$ \textbf{then} \\
\quad \quad \quad\textbf{if} $t = 1$ \textbf{then}\\
\STATE \quad \quad \quad\quad ${\bf{W}}_{\Omega _1^{\left( t \right)}}^{\left( t \right)}{\rm{ = }}{\bf{0}}$;\\
\vspace{0.25em}
\quad \quad\quad\textbf{else if} ${2 \le {\kern 1pt} {\kern 1pt} t \le T_b}$ \textbf{then}\\
\vspace{0.25em}
\STATE \quad \quad \quad\quad${\bf{W}}_{\Omega _1^{\left( t \right)}}^{\left( t \right)}$ $\leftarrow $ using (14);\\
\vspace{0.25em}
\quad \quad\quad\textbf{else}
\vspace{0.25em}
\STATE \quad \quad\quad $\quad{\bf{W}}_{\Omega _1^{\left( t \right)}}^{\left( t \right)}$ $\leftarrow $ using (16);\\
\vspace{0.25em}
\quad\quad \quad\textbf{end if}\\
\vspace{0.25em}
\quad\quad\textbf{else}
\STATE \quad \quad\quad ${\bf{W}}_{\Omega _i^{\left( t \right)}}^{\left( t \right)}$ $\leftarrow $ using (19);\\
\quad \quad\textbf{end if}\\
\STATE \quad \quad Compute ${{\bf{R}}_{{\Omega _i^{\left( t \right)}}}^{\left( t \right)}}$.\\

\STATE \quad \quad Update ${{ {\hat{\mathbf{x}}}}_{\Omega _i^{\left( t \right)}}^T}$ using (9).\\

\quad \textbf{end for}\\
\quad \textbf{while} $t <= T_b$ \textbf{do}\\
\STATE  \quad \quad Obtain ${\Omega ^{\left( t + 1\right)}}$ using (10) and (11).
\STATE \quad \quad ${\hat {\bf{x}}_{{\Omega ^{\left( {t} \right)}}/{\Omega ^{\left( t + 1\right)}}}} = {\bf{0}}$.\\

\quad \textbf{end while}\\
\textbf{end for}\\
\textbf{$\bullet$ Step 3} (\emph{Obtain The Estimated Candidate Set})\\
\STATE Use the method in \textbf{Algorithm 1}.\\

\textbf{$\bullet$ Step 4} (\emph{Obtain Data Symbols})\\
\STATE Use the method in \textbf{Algorithm 1}.\\
\end{algorithmic}
\end{algorithm}

\subsection{The Proposed CR-EBCD Algorithm}
In the above two BCD based algorithms, the process of calculating ${{\bf{R}}_k^{\left( t \right)}}$ at each iteration dominates the total computational complexity, since it needs to perform a large-scale matrix multiplication. However, at the arbitrary $t$-th iteration, a large amount of calculations for obtaining ${{\bf{R}}_k^{\left( t \right)}}$ are futile. For instance, when solving the third subproblem for acquiring ${{\bf{R}}_3^{\left( t \right)}}$, the result of ${{\bf{g}}_1}{{\hat{\mathbf{x}}}_{1,\left( {t} \right)}^T}$ is requisite. In fact, ${{\bf{g}}_1}{{\hat{\mathbf{x}}}_{1,\left( {t} \right)}^T}$ has been calculated when obtaining ${{\bf{R}}_2^{\left( t \right)}}$. Thus, the process of calculating ${{\bf{g}}_1}{{\hat{\mathbf{x}}}_{1,\left( {t} \right)}^T}$ can be regarded as computational wastage. Against this background, the CR-EBCD algorithm is developed to avoid such unnecessary computational overhead.

\begin{table*}
\centering
\small
    \caption{A list of important variables}
    \centering
    \begin{tabular}{c|c}
    \hline
     \hline
       \textbf{ Variables}  &  \textbf{ Descriptions}  \\
        \hline
         \hline
       $\Omega^{\left( {t} \right)}$  &  the candidate set at the $t$-th iteration \\
        \hline
       $\Omega_n^{\left( {t} \right)}$  &  the $n$-th user in $\Omega^{\left( {t} \right)}$ \\
       \hline
        $T_b$ & the number of iterations in pruning the candidate set \\
        \hline
        ${K_a^{\left( {t} \right)}}$ & the size of $\Omega^{\left( {t} \right)}$ \\
       \hline
        ${\overline K _a}$  &  the size of the final constant candidate set after $T_b$ iterations \\
         \hline
        ${{\bf{W}}_n^{\left( t\right)}}$  &  the auxiliary matrix of user $n$ at the $t$-th iteration \\
         \hline
       ${\hat{\mathbf{x}}}_{n,\left( t \right)}^T$  & the signals of user $n$ at the $t$-th iteration \\
       \hline
       ${{\bf{g}}_n}$  &  the equivalent channel coefficients of user $n$ at the $t$-th iteration \\
       \hline
    \end{tabular}
    \label{sim_para}
\end{table*}

For simplicity, we introduce an auxiliary matrix ${{\bf{W}}_{{\Omega _i^{\left( t \right)}}}^{\left( t \right)}} \buildrel \Delta \over = {\sum\nolimits_{l \in {\Omega ^{\left( t \right)}},l \ne \Omega _i^{\left( t \right)}}} {{{\bf{g}}_l}{\hat{\mathbf{x}}}_{l,\left( {t} \right)}^T}$. In addition, ${{\bf{W}}_{{\Omega _i^{\left( t \right)}}}^{\left( t \right)}}$ can be further categorized into three different cases and represented in the following expression:
\begin{equation}
{{{\bf{W}}_{{\Omega _i^{\left( t \right)}}}^{\left( t \right)}}}= \begin{cases}{\sum\limits_{l \in \Omega _n^{\left( t \right)}} {{{\bf{g}}_l}{\hat{\bf{x}}}_{l,\left( {t - 1} \right)}^T,}} & i=1,\\{\sum\limits_{l \in \Omega _m^{\left( t \right)}} {{{\bf{g}}_l}{\hat{\bf{x}}}_{l,\left( t \right)}^T \!+ \!\sum\limits_{l \in \Omega _n^{\left( t \right)}} {{{\bf{g}}_l}{\hat{\bf{x}}}_{l,\left( {t - 1} \right)}^T,} } }&\!{2\!\le \!i \le \!K_a^{\left( t \right)}\! - \!1}, \\ {\sum\limits_{l \in \Omega _m^{\left( t \right)}} {{{\bf{g}}_l}{\hat{\bf{x}}}_{l,\left( t \right)}^T,} } & {i = K_a^{\left( t \right)}},\end{cases}
\end{equation}
where $1 \le m \le i - 1$ and $i + 1 \le n \le K_a^{\left( t \right)}$. ${\hat{\mathbf{x}}}_{l,\left( t - 1 \right)}^T$ and ${\hat{\mathbf{x}}}_{l,\left( t \right)}^T$ denote the perviously estimated signal at the $\left( t - 1 \right)$-th iteration and the newly estimated signal at the current $t$-th iteration, respectively. Then, according to the above three different cases, we will analyse the relationship between ${{\bf{W}}_i^{\left( t - 1 \right)}}$ and ${{\bf{W}}_i^{\left( t\right)}}$ in order to achieve a tremendous computational complexity reduction. For ease of understanding, some important variables have been summarized in \textbf{Table I}.

Firstly, in the case when $i = 1$, three different situations need to be considered, $t = 1$, $2 \le t \le {T_b}$ and ${T_b} + 1 \le t \le T$, respectively. First of all, when $t = 1$, ${\bf{W}}_{{\Omega _1^{\left( t \right)}}}^{\left( t \right)} = {\bf{0}}$ because of the initialization of $\left\{ {\hat {\bf{x}}}_{k,\left( 1 \right)}^T \right\}_{k = 1}^K$. Hence, no computational cost incurs under this circumstance. Subsequently, when $2 \le t \le {T_b}$, since the candidate set ${\Omega ^{\left( {t} \right)}}$ changes at each iteration due to the pruning mechanism, the relationship between ${\bf{W}}_{{\Omega _1^{\left( t \right)}}}^{\left( t \right)}$ and the previously estimated signals also varies dynamically. In this case, ${\bf{W}}_{{\Omega _1^{\left( t \right)}}}^{\left( t \right)}$ can only be acquired by
\begin{equation}
{\bf{W}}_{\Omega _1^{\left( t \right)}}^{\left( t \right)}{\rm{ = }}\sum\nolimits_{l \in \Omega _n^{\left( t \right)}} {{\bf{g}}_l}\hat {\bf{x}}_{l,\left( {t - 1} \right)}^T,
\end{equation}
where $i + 1 \le n \le K_a^{\left( t \right)}$.
Finally, when ${T_b} + 1 \le t \le T$, the candidate set ${\Omega ^{\left( {t} \right)}}$ remains unchanged. In this case, we replace ${\Omega ^{\left( {t} \right)}}$ by ${\Omega}$ for notational simplicity. Based on (13), we can establish the following relationship between the ${\left(t - 1 \right)}$-th iteration and the $t$-th iteration:
\begin{subequations}
\renewcommand{\theequation}
{\theparentequation-\arabic{equation}}
\begin{equation}
{\bf{W}}_{{\Omega _1}}^{\left( t \right)} = \sum\limits_{l \in \Omega _n^{}} {{{\bf{g}}_l}\hat {\bf{x}}_{l,\left( {t - 1} \right)}^T},
\end{equation}
\begin{equation}
{\bf{W}}_{{\Omega _{{\overline K_a}}}}^{\left( t - 1 \right)} = \sum\limits_{l \in \Omega _m^{}} {{{\bf{g}}_l}\hat {\bf{x}}_{l,\left( {t - 1} \right)}^T}.
\end{equation}
\end{subequations}

Combining (15-1) and (15-2), ${\bf{W}}_{{\Omega _1}}^{\left( t \right)}$ can be further expressed as
\begin{equation}
{\bf{W}}_{{\Omega _1}}^{\left( t \right)} = {\bf{W}}_{{\Omega _{{\overline K_a}}}}^{\left( t - 1 \right)} + {{\bf{g}}_{{\Omega _{{\overline K_a}}}}}\hat{\bf{x}}_{{\Omega _{{\overline K_a}}},\left( {t - 1} \right)}^T - {{\bf{g}}_{{\Omega _1}}}\hat {\bf{x}}_{{\Omega _1},\left( {t - 1} \right)}^T.
\end{equation}
It is obvious that ${\bf{W}}_{{\Omega _{{\overline K_a}}}}^{\left( t - 1 \right)}$ and ${{\bf{g}}_{{\Omega _1}}}\hat {\bf{x}}_{{\Omega _1},\left( {t - 1} \right)}^T$ can be reused since both of them have been computed at the ${\left( t - 1 \right)}$-th iteration. Therefore, only the calculation of ${{\bf{g}}_{{\Omega _{{\overline K_a}}}}}\hat {\bf{x}}_{{\Omega _{{\overline K_a}}},\left( {t - 1} \right)}^T$ is required. To sum up, ${\bf{W}}_{{\Omega _1^{\left( t \right)}}}^{\left( t \right)}$ can be summarized as
\begin{equation}
\begin{aligned}
&{\bf{W}}_{\Omega _1^{\left( t \right)}}^{\left( t \right)} = \\
&\begin{cases}{\bf{0}}, & t=1,\\{\sum\limits_{l \in \Omega _n^{\left( t \right)}} {{{\bf{g}}_l}{\bf{\hat x}}_{l,\left( {t - 1} \right)}^T,} }& {2 \le {\kern 1pt} {\kern 1pt} t \le {T_b}}, \\ {{\bf{W}}_{{\Omega _{{\overline K_a}}}}^{\left( {t - 1} \right)}\! + \! {{\bf{g}}_{{\Omega _{{\overline K_a}}}}}{\bf{\hat x}}_{{\Omega _{{\overline K_a}}},\left( {t - 1} \right)}^T\! - \! {{\bf{g}}_{{\Omega _1}}}{\bf{\hat x}}_{{\Omega _1},\left( {t - 1} \right)}^T}, & {{T_b} \!+ \!1\! \le\! {\kern 1pt} {\kern 1pt} t \le T}.\end{cases}
\end{aligned}
\end{equation}

Next, we discuss the case when $2 \le i \le K_a^{\left( t \right)} - 1$. Due to the constant of $\Omega^{\left( t \right)}$ at the arbitrary $t$-th iteration, we still use ${\Omega}$ to replace ${\Omega ^{\left( {t} \right)}}$ for simplicity. Here, ${\bf{W}}_{\Omega _{i - 1}}^{\left( t \right)}$ and ${\bf{W}}_{\Omega _{i}}^{\left( t \right)}$ can be expressed as
\begin{subequations}
\renewcommand{\theequation}
{\theparentequation-\arabic{equation}}
\begin{equation}
\begin{array}{l}
{\bf{W}}_{\Omega _{i - 1}}^{\left( t \right)} = \sum\limits_{l \in \Omega _{\tilde m}} {{{\bf{g}}_l}\hat {\bf{x}}_{l,\left( t \right)}^T + \sum\limits_{l \in \Omega _{\tilde n}} {{{\bf{g}}_l}\hat{\bf{x}}_{l,\left( {t - 1} \right)}^T,} } \\
\quad\quad\quad i\ne 2, 1 \le \tilde m \le i - 2,i \le \tilde n \le K_a^{\left( t \right)}.
\end{array}
\end{equation}
\begin{equation}
\begin{array}{l}
{\bf{W}}_{\Omega _i}^{\left( t \right)} = \sum\limits_{l \in \Omega _m} {{{\bf{g}}_l}\hat {\bf{x}}_{l,\left( t \right)}^T + \sum\limits_{l \in \Omega _n} {{{\bf{g}}_l}\hat {\bf{x}}_{l,\left( {t - 1} \right)}^T,} } \\
\quad\quad\quad1 \le m \le i - 1,i + 1 \le n \le K_a^{\left( t \right)}.
\end{array}
\end{equation}
\end{subequations}
It is worth noting that if $i = 2$, ${\bf{W}}_{\Omega _{1}}^{\left( t \right)}$ should be computed by (17), rather than (18-1). According to (18), the relationship between the ${\left( i - 1 \right)}$-th candidate and the $i$-th candidate in ${\Omega}$ can be formulated as
\begin{equation}
{\bf{W}}_{\Omega _i}^{\left( t \right)} = {\bf{W}}_{\Omega _{i - 1}}^{\left( t \right)} + {{\bf{g}}_{\Omega _{i - 1}}}\hat {\bf{x}}_{\Omega _{i - 1},\left( t \right)}^T - {{\bf{g}}_{\Omega _i}}\hat {\bf{x}}_{\Omega _i,\left( {t - 1} \right)}^T.
\end{equation}
Following the same analysis as (16), only computing ${{\bf{g}}_{\Omega _{i - 1}}}\hat {\bf{x}}_{\Omega _{i - 1},\left( t \right)}^T$ incurs computational cost for obtaining ${\bf{W}}_{\Omega _i}^{\left( t \right)}$, since ${\bf{W}}_{\Omega _{i - 1}}^{\left( t \right)}$ and ${{\bf{g}}_{\Omega _i}}\hat {\bf{x}}_{\Omega _i,\left( {t - 1} \right)}^T$, as previous computing results, can be reutilized directly.

Finally, when $i = K_a^{\left( t \right)}$, we replace $K_a^{\left( t \right)}$ by $K_a$ for notational simplicity. Then, the expression of ${\bf{W}}_{{\Omega _{\left( {{K_a} - 1} \right)}}}^{\left( t \right)}$ and ${\bf{W}}_{{\Omega _{{K_a}}}}^{\left( t \right)}$ can be written as
\begin{subequations}
\renewcommand{\theequation}
{\theparentequation-\arabic{equation}}
\begin{equation}
\begin{array}{l}
{\bf{W}}_{{\Omega _{\left( {{K_a} - 1} \right)}}}^{\left( t \right)} = \sum\limits_{l \in \Omega _{\tilde m}} {{{\bf{g}}_l}\hat {\bf{x}}_{l,\left( t \right)}^T + } {{\bf{g}}_{{\Omega _{{K_a}}}}}\hat {\bf{x}}_{{\Omega _{{K_a}}},\left( {t - 1} \right)}^T,\\
\quad\quad\quad \quad1 \le \tilde m \le {K_a} - 2,
\end{array}
\end{equation}
\begin{equation}
\begin{array}{l}
{\bf{W}}_{{\Omega _{{K_a}}}}^{\left( t \right)} = \sum\limits_{l \in \Omega _m} {{{\bf{g}}_l}\hat {\bf{x}}_{l,\left( t \right)}^T,} \quad 1 \le m \le {K_a} - 1.
\end{array}
\end{equation}
\end{subequations}
Combining (20-1) and (20-2), we obtain
\begin{equation}
{\bf{W}}_{{\Omega _{{K_a}}}}^{\left( t \right)} = {\bf{W}}_{{\Omega _{\left( {{K_a} - 1} \right)}}}^{\left( t \right)} + {{\bf{g}}_{{\Omega _{{K_a}-1}}}}\hat {\bf{x}}_{{\Omega _{{K_a}-1}},\left( t \right)}^T - {{\bf{g}}_{{\Omega _{{K_a}}}}}\hat{\bf{x}}_{{\Omega _{{K_a}}},\left( {t-1} \right)}^T.
\end{equation}
Similarly, to calculate ${\bf{W}}_{{\Omega _{{K_a}}}}^{\left( t \right)}$, we only need to compute ${{\bf{g}}_{{\Omega _{{K_a}-1}}}}\hat {\bf{x}}_{{\Omega _{{K_a}-1}},\left( t \right)}^T$ in (21). Additionally, it can be observed that (19) and (21) become identical when setting $i = K_a$ in (19). Therefore, these two situations can be combined into one.

The procedure of our proposed CR-EBCD algorithm is summarized in \textbf{Algorithm 2}. Following the above discussion, we can perform the data update with extremely low computational overhead. Furthermore, after iteration process, the method to obtain the estimated candidate set and transmitted data symbols is same as that of the original BCD algorithm (Step 3 and Step 4). It is worth noting that even though the majority of inactive user have been discarded in Step 2, the interference cannot be absolutely eliminated. Thus, LS method is still applied to estimate the data symbols for a more balanced MUD performance and computational complexity trade-off.

Compared with our proposed EBCD algorithm, the iteration process (Step 2) in \textbf{Algorithm 2} is refined, while other steps are the same. To be specific, if the lines 2-5 in \textbf{Algorithm 2} are removed, our proposed CR-EBCD algorithm will degenerate as the proposed EBCD algorithm. In essence, by constantly reusing the previous calculation results, our proposed CR-EBCD algorithm is able to avoid substantial repeating computations, contributing to dramatic computational cost saving without sacrificing any detection performance.

\section{COMPUTATIONAL COMPLEXITY ANALYSIS}
In this section, the computational complexity of our proposed BCD based algorithms for uplink grant-free NOMA systems is discussed. It is noted that the total number of complex multiplications is evaluated as a criterion to measure the computational complexity \cite{gf-3}.
\vspace{-1em}

\subsection{The Original BCD Algorithm}
Firstly, for the original BCD algorithm, the complexity of calculating ${{\bf{R}}_k}$ (line 2) is $NJ{{\left( {K - 1} \right)}}$. Then, the complexity of estimating ${{{\hat{\mathbf{x}}}}_k^T}$ (line 3) is $NJ + N$. Thus, the total computational cost for completing the iteration process is $TK\left(KNJ + N\right)$. After that, the complexity for obtaining the estimated support which involves the operations of $l_2$-norm and support merge is $2K$. In Step 4, the LS operation has a complexity of ${J{\left( {2Ns^2+s^3} \right)}}$, where $s$ represents the estimated user sparsity level. Consequently, the entire computational complexity of the original BCD algorithm can be approximated as follows:
\begin{equation}
\begin{aligned}
{C_{\text{Original BCD}}} =& TK\left(KNJ + N\right) + 2K + J\left(2Ns^2+s^3\right).\\
\end{aligned}
\end{equation}

\subsection{Proposed EBCD Algorithm}
With respect to our proposed EBCD algorithm, the computational complexity of updating signals at the $t$-th iteration is ${({K_a^{\left( {t} \right)}}NJ + N)}$. When the number of iterations $t$ is less than $T_b$, the operation of pruning the candidate set requires $2{K_a^{\left( {t} \right)}}$ at each iteration. Furthermore, when the iteration number exceeds ${T_b}$, the complexity analysis can refer to that in \textbf{Algorithm 1} due to the invariant candidate set. Finally, after iteration process, the complexity of Step 3 is $2{\overline K_a}$. For Step 4, the computational cost is the same as that of the original one (Algorithm 1), i.e., ${J{\left( {2Ns^2+s^3} \right)}}$. Hence, the total computational complexity of our proposed enhanced BCD algorithm can be expressed as
\begin{equation}
\begin{aligned}
&{C_{\text{EBCD}}} =\underbrace {\sum\limits_{t = 1}^{{T_b}} {\left({K_a^{\left( t \right)}}\left( K_a^{\left( t \right)}NJ + N \right)+ 2K_a^{\left( t \right)} \right)} }_{t \le {T_b}}\\
&\quad \quad + \underbrace {\left( {T - {T_b}} \right) {\overline K_a} \left( {\overline  K_aNJ + N} \right)}_{{T_b} + 1 \le t \le T}\\
&\quad \quad \quad \quad + 2{\overline K_a} + J\left(2Ns^2+s^3\right).
\end{aligned}
\end{equation}

\newcounter{TempEqCnt} 
\setcounter{TempEqCnt}{\value{equation}} 
\setcounter{equation}{26} 
\begin{figure*}[ht]
\begin{equation}
\begin{aligned}
C_{\text{CR - EBCD}} &= C_{\text{CR - EBCD}}^{\left( 1 \right)} + C_{\text{CR - EBCD}}^{\left( 2 \right)} + C_{\text{CR - EBCD}}^{\left( 3 \right)} + 2{\overline K_a} + J\left( {2N{s^2} + {s^3}} \right)\\
&= \underbrace {\left({2NJK + NK - NJ + 2K}\right)}_{t = 1}+ \underbrace {\sum\limits_{t = 2}^{{T_b}} {\left( {3K_a^{\left( t \right)}NJ + K_a^{\left( t \right)}N - 2NJ + 2K_a^{\left( t \right)}} \right)} }_{2 \le t \le {T_b}}\\
& + \underbrace {\left( {T - {T_b}} \right)\left( {2NJ + N} \right){\overline K_a}}_{{T_b} + 1 \le t \le T} + 2{\overline K_a} + J\left( {2N{s^2} + {s^3}} \right).
\end{aligned}
\end{equation}
\hrulefill
\end{figure*}
\setcounter{equation}{\value{TempEqCnt}}

\begin{table*}
\caption{Computational Complexity Comparison Among Three BCD based Algorithms}
\newcommand{\tabincell}[2]{\begin{tabular}{@{}#1@{}}#2\end{tabular}}
\centering
\begin{tabular}{c|c}
\hline
 \hline
   \textbf{Algorithm}  &  \textbf{Number of complex multiplications}  \\
    \hline
     \hline
    Original BCD & $TK\left(KNJ + N\right) + 2K + J\left(2Ns^2+s^3\right)$ \\
    \hline
    Proposed EBCD & \tabincell{c}{$\sum\limits_{t = 1}^{{T_b}} {\left( {K_a^{\left( t \right)}\left( {K_a^{\left( t \right)}NJ + N} \right) + 2K_a^{\left( t \right)}} \right)}$\\
     $+ \left( {T - {T_b}} \right){\overline K _a}\left( {{{\overline K }_a}NJ + N} \right) + 2{\overline K _a} + J\left( {2N{s^2} + {s^3}} \right)$}\\
   \hline
    Proposed CR-EBCD  &  \tabincell{c}{$2NJK + NK - NJ + 2K $\\
    $+ \sum\limits_{t = 2}^{{T_b}} {\left( {3K_a^{\left( t \right)}NJ + K_a^{\left( t \right)}N - 2NJ + 2K_a^{\left( t \right)}} \right)}$\\
      $+ \left( {T - {T_b}} \right){\overline K _a}\left( {2NJ + N} \right) + 2{\overline K _a} + J\left( {2N{s^2} + {s^3}} \right)$} \\
     \hline
\end{tabular}
\end{table*}

\subsection{Proposed CR-EBCD Algorithm} 
We discuss the computational complexity of our proposed CR-EBCD algorithm in three distinct cases, $t=1$, ${2 \le {\kern 1pt} {\kern 1pt} t \le {T_b}}$ and ${{T_b} + 1 \le {\kern 1pt} {\kern 1pt} t \le T}$, respectively.

Firstly, we consider the situation when $t=1$. In this case, when estimating the first user's signals, calculating ${\bf{W}}_{\Omega _1^{( 1 )}}^{( 1 )}$ incurs no computational cost due to the initialization. Thus, the complexity of obtaining $\hat {\bf{x}}_{1,\left( 1 \right)}^T$ only comes from (9), which is $({NJ + N})$. Except for the first user, we update all the remaining users' signals sequentially based on (19) and (9), requiring $ {\left( {2NJ + N} \right)\left( {K - 1} \right)}$ in total. Furthermore, $2{K}$ complex multiplications are needed for pruning the candidate set. Thus, when $t=1$, the computational complexity can be represented as
\begin{equation}
\begin{aligned}
C_{\text{CR - EBCD}}^{\left( 1 \right)} &= NJ + N + \left( {2NJ + N} \right)\left( {K - 1} \right) + 2K\\
&= 2NJK + NK - NJ + 2K.
\end{aligned}
\end{equation}

As for the second case with ${2 \le {\kern 1pt} {\kern 1pt} t \le {T_b}}$, when $i = 1$, (14) is employed to compute ${\bf{W}}_{{\Omega _1^{\left( t \right)}}}^{\left( t \right)}$, which requires $({K_a^{\left( {t} \right)}-1})NJ$ complex multiplications. while updating ${\hat{\mathbf{x}}}_{1,\left( {t} \right)}^T$ using (9), the complexity for acquiring ${\hat{\mathbf{x}}}_{1,\left( {t} \right)}^T$ involves ${K_a^{\left( {t} \right)}}NJ+N$ multiplications. Similar to the case when $t=1$, the complexity for estimating all the remaining users' signals is $({K_a^{\left( {t} \right)}-1})\left( {2NJ + N} \right)$. Meanwhile, at each iteration, extra $2{K_a^{\left( {t} \right)}}$ multiplications are needed for pruning the candidate set. As a result, the total computational complexity from $t=2$ to $t={T_b}$ is given by
\begin{equation}
\begin{aligned}
&C_{\text{CR - EBCD}}^{\left( 2 \right)}\\
&= \sum\limits_{t = 2}^{{T_b}} {\left( {\left( {K_a^{\left( t \right)}NJ + N + ( {K_a^{\left( t \right)} - 1})\left( {2NJ + N} \right)} \right) + 2K_a^{\left( t \right)}} \right)} \\
& = \sum\limits_{t = 2}^{{T_b}} {\left( {3K_a^{\left( t \right)}NJ + K_a^{\left( t \right)}N - 2NJ + 2K_a^{\left( t \right)}} \right)}.
\end{aligned}
\end{equation}

For the last case, we consider ${{T_b} + 1 \le {\kern 1pt} {\kern 1pt} t \le T}$. Since the candidate set remains unchanged during this period, (16) and (19) can be exploited to calculate ${\bf{W}}_{{\Omega _1^{\left( t \right)}}}^{\left( t \right)}$ and ${\bf{W}}_{{\Omega _i^{\left( t \right)}}}^{\left( t \right)}, i \ne 1$, respectively. Furthermore, the procedure of pruning the candidate set is discarded. Hence, the computational complexity in this case only stems from lines 6-7 in \textbf{Algorithm 2}, which can be expressed by
\begin{equation}\tag{26}
C_{\text{CR - EBCD}}^{\left( 3 \right)} = \left( {2NJ + N} \right){\overline K_a}\left( {T - {T_b}} \right).
\end{equation}
As a consequence, the entire computational complexity of our proposed CR-EBCD algorithm can be computed according to (27).

To show the complexity comparison more explicitly, the computational complexities of three BCD based algorithms have been summarized into \textbf{Table II}.

\section{Numerical Simulations}
In this section, we evaluate the detection performances of our proposed BCD-based algorithms in uplink grant-free NOMA systems. Since our proposed EBCD algorithm and CR-EBCD algorithm achieve the same SER performance, we only show the performance of CR-EBCD algorithm in the following simulations. Moreover, to better validate the performance of our proposed BCD-based algorithms, oracle BCD algorithm and oracle CR-EBCD algorithm are adopted, both of which know the exact user sparsity before MUD.

For performance benchmarking, we consider the classical SP \cite{CS-4}, the spatial-temporal structure enhanced adaptive SP (STS-ASP) \cite{gf-5}, the TA-BSASP \cite{fw-2} and the sparsity adaptive block gradient pursuit (SA-BGP) \cite{LowCo-2} as baseline algorithms. In particular, it is assumed that the classical SP knows the exact user sparsity in advance while other existing algorithms can identify the active users adaptively. Additionally, as an upper bound of the MUD performance, the oracle LS method is employed, where the perfect knowledge of the active users is assumed at the BS.

\begin{figure*}[!t]
\centering
\includegraphics[width=155mm]{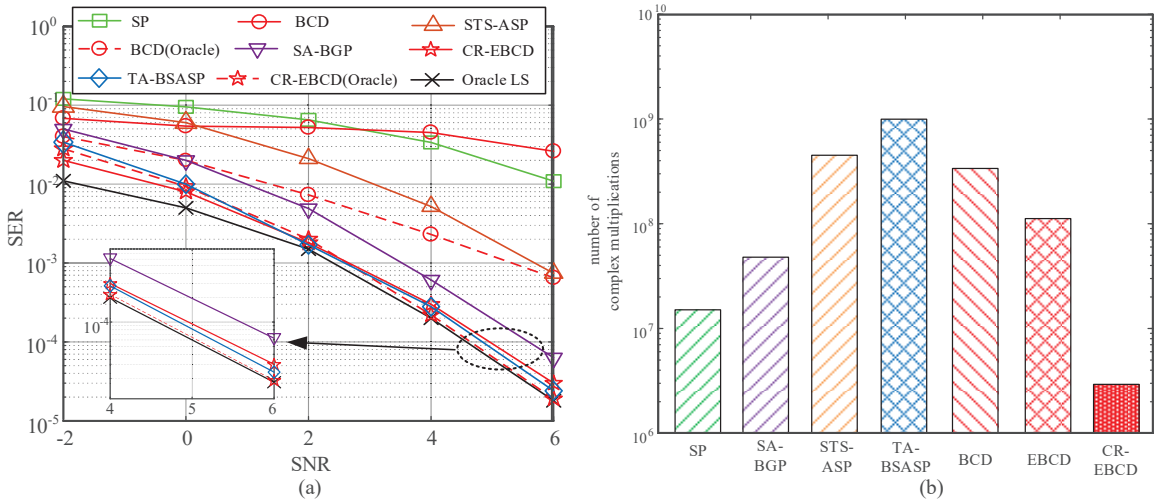}
\caption{(a) SER performance comparison of different algorithms against SNR; (b) Computational complexity comparison of different algorithms.}
\end{figure*}

\subsection{Parameters Setting}
In our simulations, the spreading sequences for each user are generated from random Gaussian sequences with the length $N = 100$, which also equals to the number of subcarriers. The total potential users is set as $K = 200$ and the actual user sparsity level varies from 18 to 20 randomly in each frame. Each frame consists of $J = 7$ continuous symbol durations according to the LTE-Advanced standard \cite{gf-8}. Furthermore, QPSK modulation is adopted. The channel coefficients in all simulations follows ${h_{n,k}} \sim {\cal C}{\cal N}\left( {0,1} \right)$.

For all BCD-based algorithms, the equivalent tuning factor $\lambda _0$ is set as 0.7 and the maximum number of iterations is equal to 12. In addition, we set the threshold $V_{th}$ to 1.31, 1.05, 0.97, 0.86, and 0.73 when signal-to-noise ratio (SNR) equals to $-2$ dB, 0 dB, 2 dB, 4 dB and 6 dB, respectively.

Finally, in Figs. 2, 5, 6 and 7, the parameter ${\psi ^{\left( t \right)}}$ for pruning the candidate set in the proposed CR-EBCD algorithms is set as ${\psi ^{\left( 1 \right)}} = {\psi ^{\left( 2 \right)}} = \cdot  \cdot  \cdot  = {\psi ^{\left( T_b \right)}} = 20$ and $T_b$ = 8.

\subsection{Simulation results}
Fig. 2(a) depicts the SER performance comparison of different algorithms versus SNR. From Fig. 2(a), it is evident that our proposed CR-EBCD algorithm significantly outperforms the classical SP, due to the exploitation of the structured sparsity. In addition, our proposed CR-EBCD algorithm achieves better CS-MUD performance compared with STS-ASP and SA-BGP and almost the same performance as TA-BSASP, thanks to the property of global optimization and the well-designed pruning mechanism. To be specific, since the proposed CR-EBCD algorithm goes through all the potential users' signals in the iteration process, the missing-detection and false-alarm are more likely to be avoided. Meanwhile, since a majority of interference from the inactive users when computing ${{\bf{R}}_k}$ is avoided with the aid of the pruning mechanism, the performance of our proposed CR-EBCD algorithm gets improved remarkably compared with that of the original BCD. It can be also observed that our proposed oracle CR-EBCD can approach the oracle LS performance. However, when it acquires the user sparsity level adaptively, it suffers from near 0.5 dB SER performance degradation at SNR of 6 dB. The reason is that the interference from inactive users cannot be completely removed due to lack of the accurate information about the actual number of active users. In this case, the stopping criterion sometimes loses its effectiveness, because the active users are hardly distinguishable from the inactive ones due to such interferences.

The comparison of the computational complexity among different schemes is shown in Fig. 2(b). It is apparent that all the BCD-based algorithms have lower computational cost than that of TA-BSASP and STS-ASP, since the computation of matrix inversion has been skipped when updating signals. Particularly, our proposed CR-EBCD algorithm saves more than $99.7{\rm{\% }}$ computational cost compared with that of TA-BSASP. Meanwhile, it enjoys significantly lower computational overhead compared with that of SA-BGP, whose matrix inversion operations are replaced by block gradient descend. The reason is that our proposed CR-EBCD algorithm requires neither sophisticated matrix inversions nor large-scale matrix multiplications. Furthermore, among all the BCD-based algorithms, our proposed EBCD algorithm saves about one-third of the complex multiplications compared with that of the original BCD, thanks to the operation of pruning the candidate set. More importantly, our proposed CR-EBCD algorithm enjoys about two orders of magnitude lower computational complexity than the proposed EBCD algorithm by removing redundant calculations. Therefore, our proposed CR-EBCD algorithm strikes a good balance between computational complexity and SER performance. Moreover, the ultra-low computational cost can lend itself for efficient implementation.

The influence of the final candidate set size ${\overline K_a}$ in our proposed CR-EBCD algorithm is demonstrated in Fig. 3. Apparently, a better SER performance can be achieved while the value of ${\overline K_a}$ becomes smaller. The reason is that when more inactive users are discarded from the candidate set and their corresponding signals are set to zeros, the calculation results of ${{\bf{R}}_k}$ can be more accurate. In this case, when updating signals for active users, the impact of interferences from the inactive users can be largely alleviated. In contrast, severe interferences give rise to the threshold failure and further lead to the SER performance loss. Moreover, with respect to the actual deployment, the final candidate set ${\overline K_a}$ is set around 0.2$K$ considering the maximum number of potential active users in real communications \cite{gf-7}. By doing so, a majority of inter-user interference can be eliminated and the missing detection can also be avoided to a great extent. Besides, it is worth mentioning that when ${\overline K_a} = 200$, our proposed CR-EBCD algorithm equivalently degenerates to the original BCD algorithm.

Fig. 4 shows the SER performance comparison against different $T_b$ in our proposed CR-EBCD algorithm. In this simulation, the final candidate set size ${\overline K_a}$ is equal to 40 and we assume ${\psi ^{\left( 1 \right)}} = {\psi ^{\left( 2 \right)}} = \cdot  \cdot  \cdot  = {\psi ^{\left( T_b \right)}}$. Correspondingly, when $T_b$ = 2, 4 and 8, ${\psi ^{\left( t \right)}}$ for the above three cases is equal to 80, 40 and 20, respectively. From Fig. 4, it is clear that the SER performance of our proposed CR-EBCD algorithm gradually improves with decreasing $T_b$. This is because pruning the candidate set excessively in the beginning few iterations causes some actual active users discarded, resulting in the performance degradation. Meanwhile, if $T_b$ is too small, more iterations are needed which bring about higher computational complexity. Hence, the selection of $T_b$ should consider the trade-off between the speed of pruning candidate set and computational complexity. From Fig. 4, it can be seen that discarding 0.1$K$ users from the candidate set in each iteration is appropriate in the practical applications.

\begin{figure}[t]
\centering{\includegraphics[width=78mm]{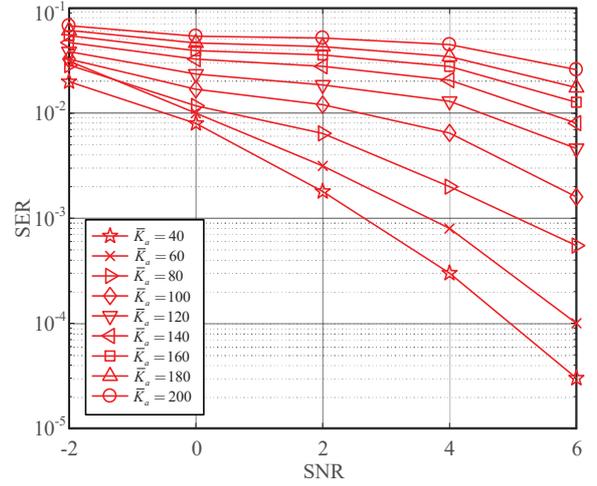}}
\caption{SER performance comparison employing our proposed CR-EBCD algorithm against different ${\overline K _a}$.}
\end{figure}

\begin{figure}[t]
\centering{\includegraphics[width=78mm]{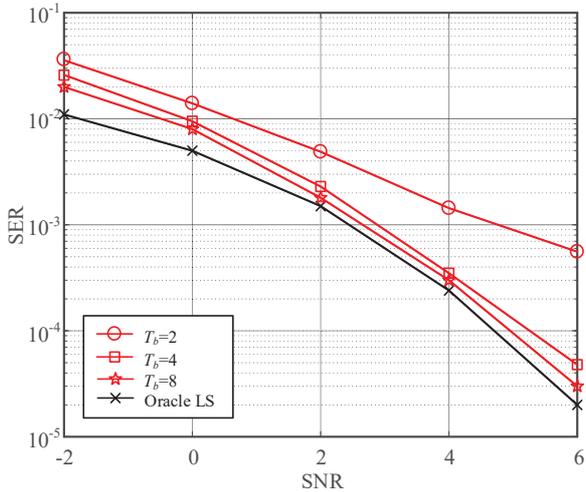}}
\caption{SER performance comparison employing our proposed CR-EBCD algorithm against different $T_b$ when the final candidate set size ${\overline K_a}$ is equal to 40.}
\end{figure}

Fig. 5 investigates the SER performance comparison of the considered algorithms against the number of consecutive time slots $J$ in a certain frame, where SNR = 4 dB. As shown by the figure, the SER performances of the SP and oracle LS algorithm remain unchanged regardless of $J$, since both of them neglect the frame-wise joint sparsity structure. Conversely, as $J$ increases, the original BCD, TA-BSASP and our proposed CR-EBCD algorithm can obtain SER performance gain, owing to the utilization of the structured sparsity. Moreover, our proposed CR-EBCD algorithm can achieve almost the same SER performance of the oracle LS method when $J \ge 7$.

\begin{figure}[t]
\centering{\includegraphics[width=78mm]{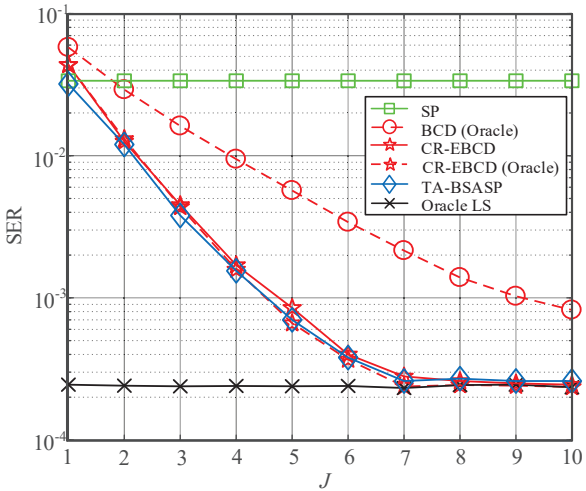}}
\caption{SER performance comparison against the number of time slots in a whole frame when SNR = 4 dB.}
\end{figure}

\begin{figure}[t]
\centering{\includegraphics[width=78mm]{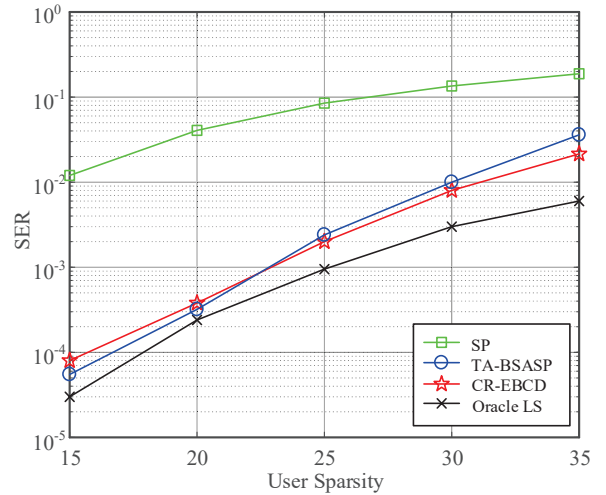}}
\caption{SER performance comparison against the active user sparsity when SNR = 4 dB.}
\end{figure}

\begin{figure}[t]
\centering{\includegraphics[width=78mm]{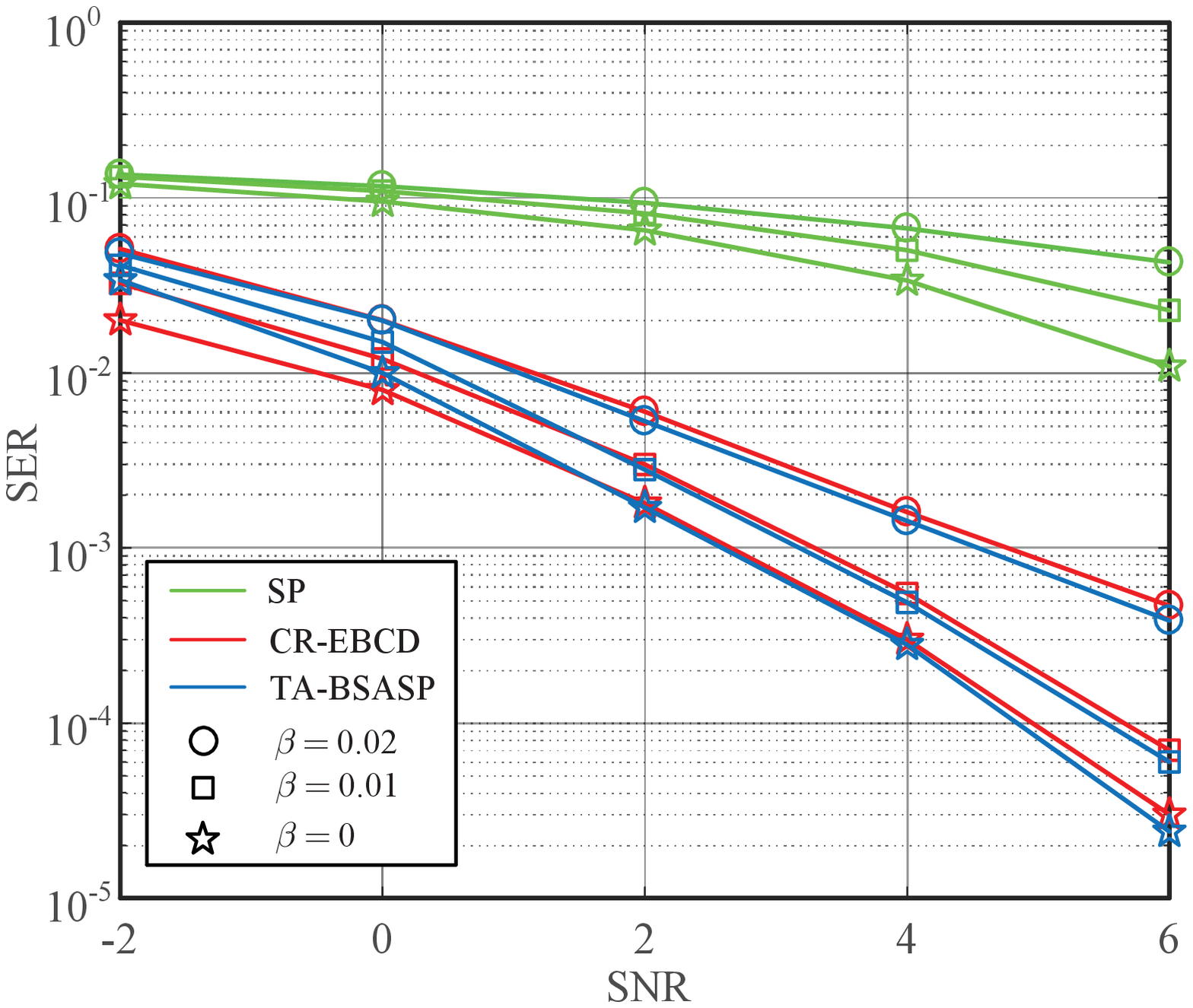}}
\caption{SER performance comparison considering time-varying channel.}
\end{figure}

Fig. 6 illustrates the impact of the user sparsity level on the SER performance of the considered algorithms when SNR = 4 dB. Simulation results indicate that as the user sparsity level rises, all the considered algorithms suffer from the SER performance degradation. Additionally, our proposed CR-EBCD algorithm outperforms the TA-BSASP algorithm when the user sparsity level is in a high region. Note that the computational cost of our proposed CR-EBCD algorithm remains the same regardless of the user sparsity, while the computational complexity of the TA-BSASP algorithm increases dramatically with the increasing actual user sparsity. Consequently, our proposed CR-EBCD algorithm is not only more tolerable to the increasing user sparsity, but also enjoys an extremely lower computational overhead compared with the TA-BSASP algorithm.

In the above simulations, similar to some existing works \cite{fw-1}-\cite{fw-6}, it is assumed that the channel information is perfectly known at the receiver side and keeps static during a data frame. However, in the practical scenarios, small channel fluctuations might occur during an entire frame, which leads to channel estimation errors. Under this circumstance, we consider a more practical scenario, where the channel for each active user varies slowly and the initial channel information ${{\bf{h}}_k^{\left( 1 \right)}}$ obtained by channel estimation will be used for the whole frame \cite{gf-4}. Thus, channel information used in the detection is no longer perfect. In this scenario, for active user $k$, the real channel gain in the $j$-th time slot can be expressed as
\setcounter{equation}{27}
\begin{equation}
\widetilde {\bf{h}}_k^{\left( j \right)} = \left( {1 - \beta } \right){{\bf{h}}_k^{\left( j - 1 \right)}} + \beta \Delta {\bf{h}}_k^{\left( j \right)},
\end{equation}
where $\beta \ll 1$ indicates the channel variation degree. Moreover, $\Delta {\bf{h}}_k^{\left( j \right)}$ represents the actual channel fluctuation of user $k$ in the $j$-th time slot and each element $\Delta {\bf{h}}_{n,k}^{\left( j \right)}$ follows ${\cal C}{\cal N}\left( {0,1} \right)$. From Fig. 7, we can observe that all the considered algorithms experience SER performance loss in the presence of channel estimation errors ($\beta = 0$ indicates time-invariant channel and perfect channel estimation). In particular, the faster the channel varies, the more the SER performance degrades. Despite of this, when $\beta = 0.02$, the SER performance of our proposed CR-EBCD algorithm can still reach $4.6 \times {\rm{1}}{{\rm{0}}^{{\rm{ - }}3}}$ when SNR = 6 dB.
\vspace{-0.3em}

\section{Conclusion}
In this paper, we have proposed two modified BCD algorithms to carry out MUD in the uplink grant-free NOMA systems. Firstly, compared with the original BCD algorithm, a novel candidate set pruning mechanism has been applied to the proposed EBCD algorithm, which significantly alleviates the interference from the inactive users. Moreover, derived from the proposed EBCD algorithm, our proposed CR-EBCD algorithm achieves substantial computational complexity reduction without sacrificing any SER performance, attributing to the elimination of all the redundant calculations in the iteration process. Then, we have carried out the computational complexity analysis of two proposed algorithms. Finally, simulation results have demonstrated that our proposed EBCD and CR-EBCD algorithms can achieve near oracle LS performance without the priori knowledge of user sparsity level. Additionally, our proposed CR-EBCD algorithm also enjoys a two orders of magnitude complexity reduction compared with the proposed EBCD algorithm, making it a feasible candidate for the practical implementation.

As a future work, it is interesting to study the case of asynchronous uplink grant-free NOMA systems. Alternatively, another promising future work is to design a multi-user detector, which enables the BS to execute the multi-user detection and synchronization jointly. In addition, the integration of grant-free NOMA systems and deep neural network (DNN) can be applied to achieve end-to-end communications for further performance enhancement \cite{ML-1}-\cite{ML-3}.

\end{document}